\documentclass[nofootinbib,superscriptaddress, prl]{revtex4}

\usepackage{amsfonts,amssymb,amsthm,bbm,amsmath, stackrel}
\usepackage{array}

\usepackage{color,psfrag}
\usepackage[dvips]{graphicx}

\newcommand{\C}{{\mathbb C}}
\newcommand{\N}{{\mathbb N}}

\newcommand{\cA}{{\mathcal A}}
\newcommand{\cB}{{\mathcal B}}

\newcommand{\cR}{{\mathcal R}}
\newcommand{\cO}{{\mathcal O}}

\newcommand{\cD}{{\mathcal D}}
\newcommand{\cC}{{\mathcal C}}

\newcommand{\cU}{{\mathcal U}}
\newcommand{\SU}{\mathrm{SU}}

\newcommand{\be}{\begin{equation}}
\newcommand{\ee}{\end{equation} }
\newcommand{\beq}{\begin{eqnarray}}
\newcommand{\eeq}{\end{eqnarray}}
\newcommand{\bes}{\begin{eqnarray}}
\newcommand{\ees}{\end{eqnarray}}

\newcommand{\su}{{\mathfrak su}}

\renewcommand{\u}{{\mathfrak u}}
\renewcommand{\sl}{{\mathfrak sl}}

\newcommand{\ra}{\rangle}

\newcommand{\f}{\frac}

\def\nn{\nonumber}

\def\mone{^{{-1}}}
\def\cop{\Delta}

\def\dr{\rightarrow}
\def\ie{\textit{i.e. }}
\def\cf{\textit{cf }}
\def\UQ{{\cU_{q}(\su(2))}}
\def\ot{\otimes}
\def\one{{\bf 1}}
\def\act{\triangleright}

\def\bt{{\bf t}}

\def\tjm{{\bf t}^{j}_{m}}

\def\demi{\f{1}{2}}

\newcommand{\CG} [2] {\,_{q}\textbf{C}\begin{array} {#1} #2 \end{array}}

\newtheorem{theorem}{Theorem}[section]

\newtheorem{prop}[theorem]{Proposition}

\newtheorem{definition}[theorem]{Definition}

\begin{document}

\title{Quantum hyperbolic geometry in loop quantum gravity with cosmological constant}

\author{{\bf Ma\"it\'e Dupuis}}\email{maite.dupuis@gravity.fau.de}
\affiliation{University Erlangen-Nuremberg, Institute for Quantum Gravity, Erlangen, Germany}

\author{{\bf Florian Girelli}}\email{fgirelli@uwaterloo.ca}
\affiliation{Department of Applied Mathematics, University of Waterloo, Waterloo, Ontario, Canada}
\affiliation{University Erlangen-Nuremberg, Institute for Quantum Gravity, Erlangen, Germany}

\begin{abstract}

Loop Quantum Gravity (LQG) is an attempt to describe the quantum gravity regime. Introducing a non-zero cosmological constant $\Lambda$ in this context has been a withstanding problem. 
Other approaches, such as Chern-Simons gravity, suggest that quantum groups can be used to introduce $\Lambda$ in the game. Not much is known when defining LQG  with a quantum group. 
Tensor operators can be used to construct observables in any type of discrete quantum gauge theory with a classical/quantum gauge group. We illustrate this by constructing explicitly
 geometric observables for LQG defined with a quantum group and show for the first time that they encode  a quantized hyperbolic geometry. This is a novel argument pointing out   the usefulness of quantum groups as encoding a non-zero cosmological constant.  
We  conclude by discussing how tensor operators provide the right formalism to unlock the LQG formulation  with  a non-zero cosmological constant.
\end{abstract}

\maketitle

%%%%%%%%%%%%%%%%%
\section*{Introduction}
%%%%%%%%%%%%%%%%%%%%%%%%%%%%%

Current cosmological data  show that our universe has a positive cosmological constant $\Lambda=10^{-52}m^{-2}$. It is therefore crucial to build a theory of quantum gravity with a non-vanishing cosmological constant $\Lambda$. 
A proposal to incorporate $\Lambda\neq0$ in the  quantum gravity regime is to work with the quantum group $\UQ$ as gauge group instead of the Lie group $\SU(2)$, where the deformation parameter, $q$, is related to $\Lambda$ \cite{rovelli}. As such, $\Lambda$ is  considered as a fundamental parameter like  Newton constant $G$ \cite{rovelli}. 
 The motivation for using quantum groups comes essentially from the quantization of 3d  models \cite{turaev, taylor}, following Witten's insights \cite{witten}. 
 The path integral quantization can  be applied to 4d  models using a quantum group \cite{fk,4dmodels}.  Preliminary results point out that in the semi-classical limit, one recovers the Regge action with a cosmological constant \cite{fk,limits}.    
 However, from a canonical quantization perspective, it is not clear why a quantum group should appear. Indeed, in the presence of a cosmological constant, the kinematical space is still built from the classical group $\SU(2)$. The cosmological constant appears in the Hamiltonian constraint, and somehow it is expected that solving this constraint would make a quantum group to appear \cite{noui}. In this paper, we do not directly address this issue. Instead, we define the $\UQ$-LQG  fundamental geometric operators\footnote{Major and Smolin proposed a way to define geometric observables using  \textit{loop} variables in the quantum group case \cite{major}. 
 However Major showed later there were important issues with their construction \cite{major issue}.}  and we show how they   encode a quantized hyperbolic geometry.  That is, we give the first insight that a quantum group in the context of LQG can really encode the presence of the presence of a hyperbolic geometry induced by $\Lambda$.
 
  Such a geometric comprehension is a first step in constructing LQG with a non-zero cosmological constant and relating it with spinfoam models based on quantum groups. This is a also strong indication that the LQG kinematical space should be fully deformed.  

Our approach is based on well-known  objects which have been under-appreciated in the LQG context, the so-called \emph{tensor operators}. They
can be used to construct observables in any  discrete quantum gauge theory \cite{Rittenberg:1991tv}.  We  show here, in the context of LQG, how they allow to construct in a straightforward manner any observables for an intertwinner and hence for a spin network. 
We illustrate the construction by considering the quantization of the triangle, which would typically  appear in 3d LQG. By considering the distance, angle and area operators we show how $\UQ$ induces the notion of hyperbolic geometry. We comment then on the extension of these results to the 4d case. In the concluding section, we discuss why tensor operators will be the relevant structure to understand the appearance of a quantum group in LQG with $\Lambda\neq0$, {\it  at the kinematical level}.  To start, let us recall the basic properties of $\UQ$ and the tensor operator definition in this case.

%%%%%%%%%%%%%%%%%%%%%%%%%%%%%
\section{$\UQ$ and tensor operators}
%%%%%%%%%%%%%%%%%%%%%%%%%%%%%

We quickly review  the essential features of $\UQ$ to fix the notations, with $q$ real\footnote{4d Lorentzian models spinfoam models are indeed constructed with $q$ real \cite{4dmodels}.  
We shall comment on the case $q$ root of unity in the last section.}. We refer to \cite{chari} for a full description.  The quasi-triangular Hopf algebra  $\UQ$ is  generated by the elements $J_{\pm},\,  K= q^{\f{J_{z}}{2}}$ such that 
\be \label{commutationSU2q}
K J_{\pm}K\mone=q^{\pm\demi} J_\pm, \quad [J_+, J_-]= \f{K^{2}-K^{-2}}{q^{1/2}-q^{-1/2}}.
\ee
The  
coproduct $\cop: \UQ \rightarrow \UQ \ot \UQ$ and antipode $S: \UQ \rightarrow \UQ$  are given by
\bes \label{deform sum angular}
&&\Delta K=K\otimes K, \quad \Delta J_\pm= J_\pm \otimes K+K\mone\otimes J_\pm, \nn\\ 
&&SK=K^{-1}, \quad S J_\pm=-q^{\pm 1/2} J_\pm.
\ees
The $\mathcal{R}$-matrix $\cR\in \, \UQ\ot \UQ$ encodes the quasi-triangular structure, which tells us how much  the coproduct is non-commutative. If we note $\sigma: \UQ\ot \UQ\dr \UQ\ot \UQ$ the permutation, then we have 
\beq\label{braiding}
\sigma \circ  \cop X = \cR (\cop X) \cR\mone, \; \forall X\in\UQ.
\eeq
Standard  notations are $\cR_{12}= \sum R_1\ot R_2$, $\cR_{21}= \sum R_2\ot R_1$, ... 
When $q$ is real, the representation theory of $\UQ$  is essentially the same as that of $\su(2)$ 
\cite{chari}. 
A representation $V^{j}$ is hence generated by the vectors $|j,m\ra$ with $j\in\N/2$ and $m\in \left\{ -j,..,j\right\}$.   The key-difference is that we use $q$-numbers $[x]\equiv \f{q^{x/2}-q^{-x/2}}{q^{1/2}-q^{-1/2}}$.  
\bes\label{vector transformation}
&& K\, |j,m\ra= q^{\f{m}{2}} |j,m\ra, \nn\\
&& J_{\pm}\, |j,m\ra = \sqrt{[ j \mp m][j\pm m +1]} \; |{j}, {m\pm1}\ra.
\ees 
The adjoint action of $\UQ$ on some operator $\cO$ is 
\be
J_{\pm} \triangleright \cO= {J_{\pm}} \cO K\mone-q^{\pm1/2}K\mone \cO {J_{\pm}}, \,\, K\act \cO = K\cO K\mone.\nn
\ee
The general definition of a \emph{tensor operator} in the case of a quasi-triangular Hopf algebra $\cA$ is given in \cite{Rittenberg:1991tv}. We review  this formalism focusing on the case $\cA= \UQ$, with $q$ real. 
The standard case of $\su(2)$ can be recovered by performing the limit $q\dr 1$.

\begin{definition} \cite{Rittenberg:1991tv}  
Let $V$ and $W$ be some $\UQ$ modules and $L(W)$ the set of linear maps on $W$.  A tensor operator $\bt$ is defined as the intertwinning linear map
\beq\label{defTO}
\begin{array}{rcl} \bt : V&\dr& L(W) \\
x&\dr& \bt(x)
\end{array}
\eeq
If we take $V\equiv V^{j}$ the representation of rank $j$ spanned by vectors $|j,m\ra$, then we note $\bt(|j,m\ra)\equiv \bt^{j}_{m}$.
\end{definition}
The fact that we have an intertwining map puts stringent constraints on the way $\bt^{j}_{m}$ transforms under $\UQ$. As an operator, $\bt^{j}_{m}$ transforms under the adjoint action but as an intertwining map it also transforms like a vector $|j,m\ra$.  Hence we have the equivariance property
\bes\label{TOtransformation}
K\act \tjm&=& K\tjm K\mone = q^{m} \tjm \nn\\
J_{\pm}\act \tjm&=& J_\pm \; \tjm \; K\mone - q^{\pm\demi} K\mone \; \tjm \; J_\pm \nn\\
& =& \sqrt{[ j \mp m][j\pm m +1]} \; \bt^{j}_{m\pm1}
\ees
The equivariance property implies the following well-known theorem. 
\begin{theorem} (Wigner-Eckart) \cite{BiedenharnBook}\label{WE}
The matrix elements   $\langle j_{1},m_{1}| \bt^{j}_{m} | j_{2},m_{2}\rangle$  are proportional to the $q$-deformed Clebsch-Gordan (CG) coefficients.  The constant of proportionality  $N_{j}(j_{1},j_{2}) $ is a function of $j_{1}, \,j_{2}$ and $j$ only.
\end{theorem}
\vspace{-.8cm}
\beq
\langle j_{1},m_{1}| \bt^{j}_{m} | j_{2},m_{2}\rangle = N_{j}(j_{1},j_{2}) \CG{c@{}c@{}c} {j& j_2& j_{1} \\ m& m_2 & m_{1}}
\eeq
Just as we can decompose the tensor product of vectors into vectors thanks to the CG coefficients, we can decompose the \textit{product} of tensor operators into tensor operators, using the CG coefficients.
\begin{prop} \label{productTOth} \cite{Rittenberg:1991tv} 
The product of tensor operators is still a tensor operator.  
\end{prop}

We can  use the Clebsch-Gordan coefficients to combine products of tensor operators.
\beq\label{productTO2}
\bt^{j}_{m}  = \sum_{m_{1}m_{2}} \CG{c@{}c@{}c}{j_{1}& j_2& j \\ m_{1}& m_2 & m}\bt^{j_{1}}_{m_{1}}\bt^{j_{2}}_{m_{2}}.
\eeq
This will be important to construct invariant operators under the adjoint action, by projecting on the trivial representation $j=0=m$. 

The \textit{tensor product} of tensor operators is more complicated to construct in the quantum group case. Indeed, if $\bt$ is a tensor operator then $\, ^{(1)}\bt=\bt \ot \one$ is a tensor operator, but $\one\ot \bt$ is in general \textit{not} a tensor operator (it is however a tensor operator if $q=1$, \ie for $\su(2)$). The reason is that $\one\ot \bt$ can be obtained from $\bt \ot \one$ using the permutation $\psi$. However if the coproduct is not co-commutative, the permutation is not an intertwining map. The solution  is then to use the $\cR$-matrix to construct an intertwining map from the permutation \cite{chari}.  

\begin{prop} \label{tensor prod}\cite{Rittenberg:1991tv}  
If  $\bt$  is a tensor operator of rank $j$ then $\, ^{(1)}\bt=\bt \ot \one$ and  $\, ^{(2)}\bt=\sigma_{\cR}(\bt \ot \one) \sigma_{\cR}\mone= \cR_{21} (\one\ot \bt)\cR_{21}\mone$  are tensor operators of rank $j$, where $ \sigma_{\cR}= \sigma\circ \cR$    is the deformed permutation.
\end{prop} 
The construction can be extended to an arbitrary number $N$ of tensor products. Starting from a given $\bt$ of rank $j$, we can build $N$ tensor operators of rank $j$ using consecutive deformed permutations. For all $i \in \{1, \cdots, N\}$,  
\beq
 ^{(i)}\bt=( \cR _{i {i-1}}  .. \cR _{i{1}}(\one \ot... \ot \bt)  \cR_{i {1}}\mone..  \cR_{i {i-1}}\mone) \ot \one\ot ..\one\nn.
\eeq
Contrary to the $q=1$ case, the operator $\,^{(i)}\bt^{j}_{m}$ does not act only on the $i^{th}$ Hilbert space $V^{j_{i}}$. It is acting non-trivially on all the Hilbert spaces  $V^{j_{k}}$ with $k\leq i$.

\medskip

Now that we have recalled the general theory of tensor operators, we can focus on their specific realization. For this we have to solve \eqref{TOtransformation}.  Just as for representations, the fundamental building blocks  are operators of rank  $1/2$, the \emph{spinor} operators $\bt^\demi$. Using the Jordan-Schwinger realization of $\UQ$, these spinor operators can be realized in terms of $q$-harmonic oscillators \cite{BiedenharnBook}. For our current purpose, we are interested in the vector operator $\bt^1$. They can be realized in terms of either the  $q$-harmonic oscillators or the $\UQ$ generators.
\bes
&&\bt^{1}_{\pm1}=
\mp \f{q^{\f{J_z}{2}}}{\sqrt{[2]}}J_\pm, 
\,\,\,
\bt^{1}_0=
\f1{{[2]}}(q^{-1/2}J_+J_--q^{1/2}J_-J_+). \nn
\ees
When $q=1$, the vector operator components $\bt^{1}_\alpha$ are proportional to the $\su(2)$ generators $J_{\alpha}$.

Any other tensor operator of rank $j$ can be built by combining spinor operators and CG coefficients thanks to Proposition 3. Then, the construction of tensor operators of rank $j$ from tensor products of a given tensor operator can be done using Proposition 4. 
Contrary to the undeformed case, the components of $\,^{(a)}\bt^{j}$ and $\,^{(b)}\bt^{j}$ will not commute in general for $a\neq b$.

\vspace{-.3cm}

%%%%%%%%%%%%%%%%%%%%%%%%%%%%%
\section{Quantum hyperbolic geometry}
%%%%%%%%%%%%%%%%%%%%%%%%%%%%%
In LQG with $\Lambda=0$, the quantization procedure leads to spin networks states, which are graphs decorated by $\su(2)$ representations $j_i$ on the edges and intertwiners  $|\iota_{j_1..j_N}\ra$ on the vertices with N legs.  The fundamental chunk of quantum space is given by the intertwinner, a vector of $V^{j_1}\ot \cdots \ot V^{j_N}$ invariant under $\su(2)$.  
To encode $\Lambda\neq0$, we replace $\su(2)$ by $\UQ$, with $q$ real and work with   $\UQ$ spin networks.  We expect then that a $\UQ$ intertwiner should describe a quantum hyperbolic  chunk of space.  Observables acting on the intertwiner space are now easy to construct. They 
are  tensor operators of rank $j=0$, since by definition they are  invariant under the  adjoint action of  $\UQ$. They can be built out from  tensor operators of rank $k$, $\,^{(i)}\bt^{k}, \, i \in \{1, \cdots, N\}$ acting in $V^{j_1}\ot \cdots \ot V^{j_N}$ combined together with the relevant CG coefficients to project on the trivial rank, following Proposition 3. {Observables are therefore the intertwiners image under the map} \eqref{defTO}.  This will be true for any  type of group, classical or quantum.

For simplicity, let us focus first on 3d (Euclidian) LQG. We take  $q= e^{-\frac{\ell_p}{R}}=e^{-\lambda} $ real, with $\ell_p$ and $R$ respectively the Planck scale and the cosmological radius.
To probe the nature of the quantum geometry encoded by a $\UQ$ intertwiner, we consider the simple case of a triangle quantum state given by the three-leg intertwiner $|\iota_{j_{b}j_{c}j_{a}}\ra$ which is
\beq
\sum_{m_{i}} \f{(-1)^{j_a+m_a}q^{-\f{m_a}2}}{\sqrt{[2j_a+1]}} \CG{c@{}c@{}c}{j_{b}& j_{c} & j_{a}\\m_{b}& m_{c}& -m_{a}} |j_{b}m_{b}, j_{c}m_{c},j_{a}m_{a} \ra. \nn
\eeq
We use the vector  operators $\,^{(i)}\bt^1$, $i=a,b,c$,  to construct the $\UQ$ generalization of  the key geometrical observables such as the  angle, length, and area operators. 
In the classical case, the angle operator between the edges $i$ and $j$ is $\,^{(i)}\vec J \cdot \, ^{(j)}\vec J$, hence the natural generalization is 
\beq
\,^{(i)}\bt^{1}\cdot \, ^{(j)} \bt^{1}\equiv -\sqrt{[3]} \CG{c@{}c@{}c} {1& 1 & 0 \\ m_{1}& m_2 & 0} \,^{(i)}\bt^{1}_{m_{1}} \,^{(j)}\bt^{1}_{m_{2}},
\eeq 
which gives again  $\,^{(i)}\vec J \cdot \, ^{(j)}\vec J$ when $q\dr 1$. The  action of $\,^{(i)}\bt^{1}\cdot \, ^{(j)} \bt^{1}$ on the triangle quantum state $|\iota_{j_{b}j_{c}j_{a}}\ra$ is diagonal. If we take  $i=b$, $j=c$, the eigenvalue  is 
\bes\nn
q\frac{\cosh\frac{\lambda}{2} \cosh ((j_a+\demi) \lambda) - \cosh ((j_b+\demi)\lambda) \cosh((j_c+\demi)\lambda))}{\sqrt{(\sinh^2((j_b+\demi)\lambda) - \sinh^2\f\lambda2)(\sinh^2((j_c+\demi)\lambda) - \sinh^2\f\lambda2)}},
\ees
where $\lambda\equiv \ell_p/R$. We  recognize  a quantization of the hyperbolic cosine law (\cf Fig 1),
\beq\label{cosinelaw}
 -\hat n_{b} \cdot \hat n_{c} 
=  \cos \theta_{a}
= \frac{-\cosh \frac{l_a}{R} + \cosh \frac{l_b}{R} \cosh \frac{l_c}{R}}{\sinh \frac{l_b}{R}\sinh \frac{l_c}{R}},
\eeq
provided the edge length is quantized as $l_i\dr (j_i+1/2)\ell_p$. %
 \begin{figure}[h] \label{triangle}
\begin{center}
 \includegraphics[scale=0.7]{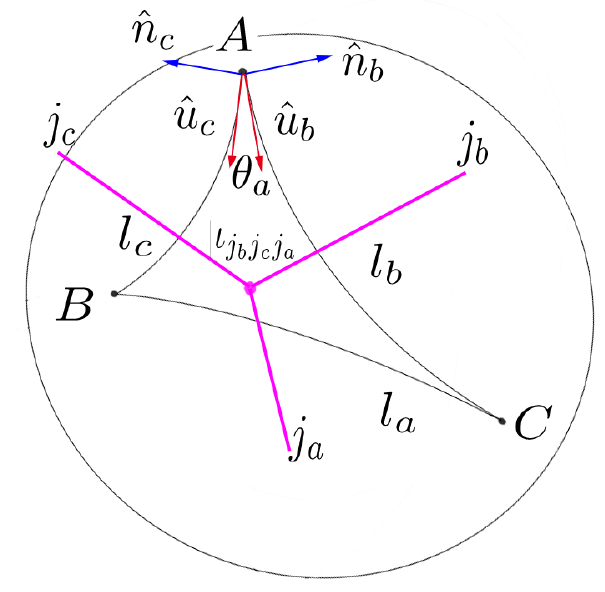}
\caption{Hyperbolic triangle in Poincar\'e disc. Normals are such that $\vec  n_{i}= \sinh \f{l_i}{R} \hat n_i$. $\hat u_i$ are the normalized tangent vectors and $\hat u_{b} \cdot \hat u_{c}=   \cos \theta_{a}$.}
\end{center}
\end{figure}
\\
We note  that the ordering factor $\cosh\frac{\lambda}{2}$ is necessary to obtain the right flat limit of the quantum cosine law \cite{val}. This factor becomes negligible in the classical limit $\ell_p\dr 0$, just as the other ordering factor $\sinh\frac{\lambda}{2}$. In a quantum theory, we usually encounter ordering ambiguities. When dealing with $\UQ$, we have further ambiguities since the tensor product of operators becomes also non-commutative. For example, we have that $\,^{(c)}\bt^{1}\cdot \, ^{(b)} \bt^{1}|\iota_{j_{b}j_{c}j_{a}}\ra = q^{-2}\,^{(b)}\bt^{1}\cdot \, ^{(c)} \bt^{1}|\iota_{j_{b}j_{c}j_{a}}\ra$. 
We finally emphasize that the square of the norm operator $\,^{(i)}\bt^1 \cdot \,^{(i)}\bt^1$ is diagonal with eigenvalue
\beq\label{norm}\f{[2j_i][2j_i+2]}{[2]}= \f{ \sinh^2((j_i+\demi)\lambda) - \sinh^2\f\lambda2}{\sinh \lambda \sinh \f{\lambda}{2}},\eeq
which is \textit{not} the square of the standard length operator, but a \textit{function of it}. Only in the limit $R\dr\infty$, this becomes the square of the length operator, $l_i^2\dr {j_i(j_i+1)\ell_p^2}$. 

Bianchi et al.  have heuristically argued   that  a minimum angle in the quantum gravity regime appears due to the presence of the cosmological constant \cite{eugenio}. This can be explicitly checked in our scheme. Setting $j_a=0$, we must have $j_b=j_c=j$, and the above quantized cosine law gives   
 \bes\label{nonzerocos}
q\frac{\cosh^2\frac{\lambda}{2} - \cosh^2 ((j+\demi)\lambda)}{{\sinh^2((j+\demi)\lambda) - \sinh^2\f\lambda2}}, %\neq 1,
\ees
which means that there is a non-zero minimum angle. When $\ell_p\dr0$ (classical limit) or $R\dr\infty$ (flat quantum limit), \eqref{nonzerocos} tends to 1, so we recover that the triangle is degenerated.  

The area of an hyperbolic triangle is given in terms of the triangle angles,
\beq
\cA= {(\pi-\theta_a-\theta_b-\theta_c)}{R^2}.
\eeq
We can express some function of the area in terms of the  triangle edges lengths  \cite{mend}. For instance $\sin^2 \f{\cA}{2R^2}$ is
\beq\label{mednykh}
 \f{\sinh(\f{s}{2R})\sinh(\f{s-l_a}{2R})\sinh(\f{s-l_b}{2R})\sinh(\f{s-l_c}{2R})}{\cosh^2\f{l_a}{2R}\cosh^2\f{l_b}{2R}\cosh^2\f{l_c}{2R}},
\eeq
where $s=\demi(l_a+l_b+l_c)$. Putting together \eqref{cosinelaw} and \eqref{mednykh}, we  have that $4 (\sin^2 \f{\cA}{2R^2} )(\cosh^2\f{l_a}{2R}\cosh^2\f{l_b}{2R}\cosh^2\f{l_c}{2R})$ is
\beq\label{areamend}
 \sinh^2\f{l_b}{R}\sinh^2\f{l_c}{R}(1-\cos^2\theta_a) =\vec n_b^2\vec n_c^2- (\vec n_b\cdot \vec n_c)^2. 
\eeq
On the other hand, for a \textit{flat} triangle, the square of the area is $\cA^2= \f{\vec n_b^2\vec n_c^2- (\vec n_b\cdot \vec n_c)^2}{4}= \f{|\vec n_b\wedge \vec n_c|^2}{4}$. 
Replacing the normal $\vec n_i$ by $\,^{(i)}\vec J$ leads to the quantization of the area \cite{area}. Note however that  due to some ordering factors, the quantized version of the last equality is not exactly true. Following \cite{area}, we generalize the quantization of $\vec n_b^2\vec n_c^2- (\vec n_b\cdot \vec n_c)^2$ to the quantum group case as
\beq
(\,^{(b)}\bt^{1}\cdot \, ^{(b)} \bt^{1})(\,^{(c)}\bt^{1}\cdot \, ^{(c)} \bt^{1})-q^{-2}(\,^{(b)}\bt^{1}\cdot \, ^{(c)} \bt^{1})^2.
\eeq 
The action of this operator on $|\iota_{j_{b}j_{c}j_{a}}\ra$ is obviously diagonal and the eigenvalue is fully expressed in terms of the quantized lengths, just as in the classical case \cite{area}. We obtain therefore   a quantization of the function of the area given by \eqref{areamend}, just as we got a quantization of a function of the length considering the norm of the vector operator.  

The tensor operator formalism can be obviously extended to the 4d setting.  The simplest geometry to consider is that of a quantum tetrahedron, given in terms of a four-leg $\UQ$ intertwiner $|\iota_{j_{1}j_{2}j_{3}j_4}\ra$.  The angle operator $\,^{(i)}\bt^{1}\cdot \, ^{(j)} \bt^{1}$ describes now the  quantization of the dihedral angle, since the vector operator will be interpreted as the normal to the face. The norm of the vector operator will be interpreted as a \textit{function} of the area. The (squared) area is therefore quantized with  eigenvalues $(j_i+\demi)\ell_p^2$, since $q=e^{-{\ell_p^2}/{R^2}}$ in the 4d case. The (square of the) volume operator  in the classical case is built from $(\,^{(i)}J\wedge \,^{(j)}J)\cdot \,^{(k)}J$, which is
$${3} \sum_{m_i,\alpha_i}{\bf C}\begin{array}{c@{}c@{}c} 1& 1 & 0 \\ \alpha_2& \alpha_1 & 0 \end{array}{\bf C}\begin{array}{c@{}c@{}c} 1& 1 & 1 \\ m_{1}& m_2 & \alpha_2\end{array} \,^{(i)}J_{m_{1}} \,^{(j)}J_{m_{2}} \,^{(k)}J_{\alpha_{1}}. $$
To generalize this to the quantum group case, we can replace the $\su(2)$ vector operators $\,^{(i)}\vec J$ by the $\UQ$ vector operators $\,^{(i)}\bt^{1}$ and use the relevant CG coefficients. We expect in this case to recover a \textit{function} of the volume of the hyperbolic tetrahedron.

%%%%%%%%%%%%%%%%%%%%%%%%%%%%%
\section*{Outlook}
%%%%%%%%%%%%%%%%%%%%%%%%%%%%%
In 3d quantum  gravity,  it is well known that the cosmological constant appears through a quantum group structure \cite{witten}. Not much is known from the LQG approach. We have  shown here, in the case of $q$ real, how the standard geometric operators of LQG are generalized to the quantum group case and characterize a quantized hyperbolic geometry. This shows explicitly that a quantum gauge group in LQG encodes   a non-vanishing cosmological constant and that the kinematical space should be deformed.
 Moreover, we  recovered that the quantum spatial  geometry is discrete and that there is a notion of minimum angle. This could  lead to potential phenomenological evidences \cite{pheno}. Tensor operators have been the key objects for this generalization. 

We have treated the case $q$ real since  the $\UQ$ representation theory is easy and it is also the relevant case to discuss the 4d quantum gravity models. Indeed current 4d Lorentzian spinfoam models are built with $q$ real \cite{4dmodels}.  There is no 4d Lorentzian model with $q$ root of unity since the $\cU_q(\sl(2,\C))$ representation theory with $q$ complex is not well understood. In the 3d Euclidian case, with $q$ root of unity, the tensor operator construction becomes potentially more complicated due to the nature of the $\UQ$ representation theory \cite{chari}. However it is quite likely that our results can then be extended  to this case and lead to quantum spherical geometries. 

We expect that the use of  tensor operators  in LQG will provide us new routes to understand how to derive LQG when the cosmological constant is not zero. First the $U(N)$ formalism,  generating all observables for an intertwiner \cite{HO},  can be extended to the quantum group case in a direct manner \cite{un}. The standard $U(N)$ formalism was used recently to rewrite the 3d LQG Hamiltonian constraint and to solve it to obtain the Ponzano-Regge spinfoam model \cite{etera}. We can use the $\cU_{q}(\u(N))$ formalism to generalize the Hamiltonian constraint to the presence of cosmological constant and relate it to the Turaev-Viro spinfoam amplitude \cite{hamil}. Finally, a nice feature of the $U(N)$ formalism is the geometrical interpretation, through twisted geometries \cite{twisted}.  Tensor operators  provide guidance on the identification of the nature of the classical variables defining {\it deformed} twisted geometries and  the phase space structure relevant to LQG with a cosmological constant \cite{phase}.

%%%%%%%%%%%%%%%%%%%%%%%%%%%%%

\end{document}